\documentclass[amsmath,showpacs,nofootinbib,11pt,superscriptaddress]{revtex4-2}
\usepackage{graphicx}
\usepackage{dcolumn}
\usepackage{bm}
\usepackage{subcaption}
\usepackage{color} 
\usepackage{slashed}
\usepackage{float}
\usepackage{amsfonts}
\usepackage{multirow}
\usepackage{lineno}
\usepackage[colorlinks=true,linkcolor=blue,citecolor=blue,urlcolor=blue]{hyperref}

\begin{document}
\newcommand{\hs}{\hspace*{0.5cm}}
\newcommand{\vs}{\vspace*{0.5cm}}
\newcommand{\be}{\begin{equation}}
\newcommand{\ee}{\end{equation}}
\newcommand{\bea}{\begin{eqnarray}}
\newcommand{\eea}{\end{eqnarray}}
\newcommand{\ben}{\begin{enumerate}}
\newcommand{\een}{\end{enumerate}}
\newcommand{\bde}{\begin{widetext}}
\newcommand{\ede}{\end{widetext}}
\newcommand{\nn}{\nonumber}
\newcommand{\crn}{\nonumber \\}
\newcommand{\Tr}{\mathrm{Tr}}
\newcommand{\non}{\nonumber}
\newcommand{\noi}{\noindent}
\newcommand{\al}{\alpha}
\newcommand{\la}{\lambda}
\newcommand{\bet}{\beta}
\newcommand{\ga}{\gamma}
\newcommand{\va}{\varphi}
\newcommand{\om}{\omega}
\newcommand{\pa}{\partial}
\newcommand{\+}{\dagger}
\newcommand{\fr}{\frac}
\newcommand{\bc}{\begin{center}}
\newcommand{\ec}{\end{center}}
\newcommand{\Ga}{\Gamma}
\newcommand{\de}{\delta}
\newcommand{\De}{\Delta}
\newcommand{\ep}{\epsilon}
\newcommand{\varep}{\varepsilon}
\newcommand{\ka}{\kappa}
\newcommand{\La}{\Lambda}
\newcommand{\si}{\sigma}
\newcommand{\Si}{\Sigma}
\newcommand{\ta}{\tau}
\newcommand{\up}{\upsilon}
\newcommand{\Up}{\Upsilon}
\newcommand{\ze}{\zeta}
\newcommand{\ps}{\psi}
\newcommand{\Ps}{\Psi}
\newcommand{\ph}{\phi}
\newcommand{\vph}{\varphi}
\newcommand{\Ph}{\Phi}
\newcommand{\Om}{\Omega}
\newcommand{\AdrHEPC}{Phenikaa Institute for Advanced Study, Phenikaa University, Nguyen Trac, Duong Noi, Hanoi 100000, Vietnam}
\newcommand{\AdrH}{Institute of Physics, Vietnam Academy of Science and Technology, 10 Dao Tan, Giang Vo, Hanoi 100000, Vietnam}

\title{Constraints from the SM-like Higgs boson in a flavor-dependent $U(1)$ extension of the Standard Model}
 
\author{Duong Van Loi}
\email{loi.duongvan@phenikaa-uni.edu.vn (corresponding author)}
\affiliation{\AdrHEPC} 
\author{N. T. Duy}
\email{ntduy@iop.vast.vn}
\affiliation{\AdrH} 

\date{\today}

\begin{abstract}
This work presents a phenomenological study of the Standard Model-like Higgs boson $H$ in a flavor-dependent $U(1)_X$ extension of the Standard Model, where the $X$ charge is assigned according to fermion flavors. In particular, we analyze the interactions of $H$ with Standard Model particles as well as with new charged scalar bosons. In addition, the parameter $\ka_{\ga}$, which characterizes the $H\to \ga\ga$ decay at the one-loop level, is investigated. The results show that the model predicts $\ka_{\ga}$ values consistent with the ATLAS and CMS constraints at the $1\sigma$ level, while satisfying several bounds derived from flavor, collider and dark matter studies. 
\end{abstract}

\maketitle	

\section{Introduction}
Neutrino masses and dark matter (DM) are two of the most challenging problems of the Standard Model (SM), which call for extensions of the SM. In addition, the SM lacks a natural explanation for several fundamental questions, such as why there are three generations of fermions, why fermion masses exhibit strong hierarchies, and the origin of several anomalies observed in quark and lepton flavor-violating processes. Finding a Beyond Standard Model (BSM) framework that can simultaneously address these puzzles is one of the main goals in particle physics. In fact, many BSM models extending the gauge symmetry or the Higgs sector of the SM have been proposed~\cite{Huong:2018ytz,Duy:2022qhy,Huong:2019vej,VanDong:2015ifg,VanLoi:2020xcq,Van_Loi_2020,Thu:2023xai,Huong:2020csh,NguyenTuan:2020xls,Dinh:2019jdg,Duy:2025jbw,VanDong:2024lry,VanLoi:2024ptt,Duy:2024txy,VanLoi:2025fmy,VanLoi:2023kgl}.
	
Recently, the authors in Ref.~\cite{VanLoi:2024ptt} proposed a BSM model based on a flavor-dependent $U(1)_X$ gauge symmetry, where the $X$ charge is defined as a function of the fermion flavor. The model provides natural and economical solutions to several open issues of the SM, including the origin of fermion generations, small neutrino masses, dark matter, and flavor anomalies. In this model, there exists a light Higgs boson $H$ whose mass naturally lies at the electroweak scale. Therefore, it can be identified with the Higgs boson with a mass of about 125~GeV discovered at the LHC~\cite{ATLAS:2012yve,CMS:2012qbp}. Importantly, $H$ mixes with two heavier Higgs bosons $H_{1,2}$ through the mixing parameters $\ep_{1,2}$. These mixings can modify the properties of $H$, such as its decay widths and production rates, compared with the SM predictions. Moreover, the ATLAS and CMS collaborations have reported results on the Higgs signal strength modifiers $\ka_i$ ($i=l,u,d,W,Z,g,\ga$) using Run II data~\cite{ParticleDataGroup:2024cfk}. These parameters quantify the deviations of the Higgs couplings from the SM predictions, defined as $\ka_i = g_i/g_i^{\mathrm{SM}}$. They therefore provide powerful probes for constraining possible new physics contributions. However, the phenomenology of the Higgs boson $H$ has not yet been studied in detail in Ref.~\cite{VanLoi:2024ptt}. Therefore, the main goal of this paper is to perform a systematic analysis of the $\ka_i$ parameters of $H$.
	
This paper is organized as follows. In Sect.~\ref{m1}, we briefly introduce the model, including the particle content, the mass spectra of the scalar and gauge bosons, and the interactions of $H$ with the SM charged fermions as well as with the charged scalar bosons. Sect.~\ref{m3} is devoted to the phenomenological analysis of $H$, including the modified parameters $\ka_i$ ($i=l,u,d,W,Z,\ga$). Finally, the main results are summarized in Sect.~\ref{m4}.
	
\section{\label{m1} Brief review of the model}
\subsection{Particle content and symmetries}
Our theoretical framework incorporates an additional Abelian gauge symmetry $U(1)_X$, resulting in the  symmetry group  $SU(3)_C\times SU(2)_L \times U(1)_Y \times U(1)_X $ ~\cite{VanLoi:2024ptt}. In this construction, the $X$ charge is assigned as a function of the generation (flavor) index $a$, the baryon and lepton numbers $B$ and $L$, and a nonzero parameter $z$, such that
\be X=3z[Bi^{a^2(a-1)}+Li^{a(a-1)}].	\ee
Due to the property that the $X$ charges of the fermion families are periodic in $a$ with a period of $4$, the number of fermion generations $N_f$ can be expressed as $N_f=4m-n$ (with $m=1,2,3,...$ and $n=0,1,2,3$). This property affects the gauge anomaly $[SU(2)_L]^2U(1)_X$, which is canceled only when $m=n=1$. Consequently, the model naturally predicts the existence of three fermion generations, $N_f = 3$, in Nature~\cite{VanLoi:2024ptt}. The fermion representations of the model are given by
	\be
	l_{1L}=\left(%
	\begin{array}{c}
		\nu_{1L} \\
		e_{1L}
	\end{array}%
	\right)\sim \left(1,2,\frac{-1}{2},3z\right), \hs 	l_{xL}=\left(%
	\begin{array}{c}
		\nu_{xL} \\
		e_{xL}
	\end{array}%
	\right)\sim \left(1,2,\frac{-1}{2},-3z\right),
	\ee
	\be
	Q_{\al L} =\left(%
	\begin{array}{c}
		u_{\al L} \\
		d_{\al L} 
	\end{array}%
	\right)\sim \left(3,2,\frac{1}{6}, z\right), \hs 	Q_{3 L} =\left(%
	\begin{array}{c}
		u_{3 L} \\
		d_{3 L} 
	\end{array}%
	\right)\sim \left(3,2,\frac{1}{6}, -z\right),
	\ee
	\be
	\nu_{R}\sim \left(1,1,0,-3z\right), \hs e_{1R}\sim \left(1,1,-1,3z\right),\hs e_{xR}\sim (1,1,-1,-3z), \hs 	N_{R}\sim (1,1,0,0), \ee 
	\be u_{\al R} \sim
	\left(3,1,\fr{2}{3},z\right), \hs d_{\al R}\sim \left(3,1,-\fr{1}{3},z\right), \hs u_{3 R} \sim
	\left(3,1,\fr{2}{3},-z\right), \hs d_{3R}\sim \left(3,1,-\fr{1}{3},-z\right),
	\ee
where $x=2,3$ and $\alpha=1,2$ denote family indices. The right-handed neutrino $\nu_R$ is introduced to cancel the $[U(1)_X]^3$ and $[\text{Gravity}]^2 U(1)_X$ anomalies. In addition, the gauge-singlet fermion $N_R$, which is neutral under the SM gauge symmetry, is included to generate light neutrino masses through the scotoseesaw mechanism.

The scalar sector of the considered model includes two doublets, $\phi$ and $\eta$, and two singlets, $\chi_1$ and $\chi_2$. Their corresponding $SU(3)_C \times SU(2)_L \times U(1)_Y \times U(1)_X$ quantum number assignments are given by
	\be
	\phi = \left(%
	\begin{array}{c}
		\phi^+  \\
		\phi^0 
	\end{array}%
	\right) \sim \left(1,2,\frac{1}{2}, 0\right),\hs 
	\eta = \left(%
	\begin{array}{c}
		\eta^{0}  \\
		\eta^- 
	\end{array}%
	\right) \sim \left(1,2,-\frac{1}{2}, 3z\right), \ee
	\be
	\chi_1 = \left(1 ,1,0, 2z\right), \hs \chi_2 = \left(1 ,1,0, 2z\right).
	\ee	
The vacuum expectation values (VEVs) are defined as
	\be \langle \phi \rangle = \fr{1}{\sqrt{2}}\left(
	\begin{array}{c} 0 \\ v \end{array}\right), \hs
	\langle \chi_1\rangle = \fr{1}{\sqrt{2}} \La_1,\hs \langle \chi_2\rangle = \fr{1}{\sqrt{2}} \La_2.\ee
	Thus, the gauge symmetry of the model is spontaneously broken, generating masses for all particles. In particular,  the $U(1)_X$ symmetry is broken by the VEVs of the singlets $\chi_1$ and $\chi_2$, which results in  a residual symmetry $R$ defined as $R=e^{i\delta X}$, where $\delta$ is the transformation parameter). Meanwhile, the doublet $\phi$ plays the same role as the SM Higgs doublet, breaking the electroweak symmetry down to the electromagnetic symmetry, i.e. $SU(2)_L\times U(1)_Y \to U(1)_Q$. The residual symmetry $R$ can be written as $R=e^{i\ka\pi X/z}=(-1)^{\ka X/z}$ for $\delta=\ka\pi/z$ ($\ka\in Z$), since $R$ preserves the VEVs of $\chi_1$ and $\chi_2$. As shown explicitly in Ref.~\cite{VanLoi:2024ptt}, the symmetry $R$ is isomorphic to the discrete group, $Z_2$ \cite{VanLoi:2024ptt}. We emphasize that all the SM fields, $\nu_R$, and $\chi_{1,2}$ transform trivially under $Z_2$ (even parity), while $N_R$ and the scalar doublets $\eta,\rho$ transform nontrivially (odd parity). In addition, the VEVs must satisfy the hierarchy $v\ll \La_1,\La_2$ in order to be consistent with the SM. 
	
\subsection{Scalar and gauge mass spectrum} 
In this section, we summarize the main results obtained from the diagonalization of the scalar potential, which were derived explicitly in Ref. \cite{VanLoi:2024ptt}. The scalar potential of the model under consideration has the form 
	\be V=V(\ph,\chi_{1,2})+V(\eta,\rho,\text{mix}),\ee
	where 
	\begin{align} 
	V(\ph,\chi_{1,2}) =& \mu_1^2 \phi^\dagger \phi + \mu_2^2 \chi_1^* \chi_1 + \mu_3^2 \chi_2^* \chi_2 + \la_1(\phi^\dagger \phi)^2 + \la_2 (\chi_1^* \chi_1)^2 + \la_3 (\chi_2^* \chi_2)^2\crn
	&+(\phi^\dagger \phi)(\la_4\chi_1^* \chi_1 + \la_5\chi_2^* \chi_2)+ \la_6 (\chi_1^* \chi_1)(\chi_2^* \chi_2)+ (\la\chi_1^3\chi_2^*+\mathrm{H.c.}),\\
	V(\eta,\rho,\text{mix}) =& \mu^2_4 \eta^\dagger \eta + \mu^2_5 \rho^\dagger \rho+\la_7(\eta^\dagger \eta)^2+\la_8(\rho^\dagger \rho)^2+\la_9(\eta^\dagger \eta)(\rho^\dagger \rho)+\la_{10}(\eta^\dagger \rho)(\rho^\dagger \eta)\crn
	& +(\ph^\dagger \ph)(\la_{11} \eta^\dagger \eta +\la_{12} \rho^\dagger \rho)+\la_{13} (\eta^\dagger\ph)(\ph^\dagger \eta)+\la_{14} (\rho^\dagger\ph)(\ph^\dagger \rho)\crn
	&+ (\chi_1^* \chi_1)(\la_{15}\eta^\dagger \eta+\la_{16}\rho^\dagger \rho)+(\chi_2^* \chi_2)(\la_{17}\eta^\dagger \eta+\la_{18}\rho^\dagger \rho)\crn
	&+[\la_{19}(\ph\eta)(\ph\rho)+\mu (\eta^\dagger \rho) \chi_2 + \mathrm{H.c.}],
	\end{align}
where the scalar couplings $\la_i$ and $\la$ are dimensionless, while the parameters $\mu_i$ and $\mu$ have mass dimension. We impose the following conditions so that the potential $V$ is bounded from below and the gauge symmetry breaking is properly induced:
	\be\mu_{1,2,3}^2<0, |\mu_1|\ll |\mu_{2,3}|, \mu_{4,5}^2>0, \la_{1,2,3,7,8}>0. \ee 
	
We expand the $Z_2$-even scalar fields $\phi,\chi_1,\chi_2$ around their VEVs and substitute them into the scalar potential. From this, we obtain the potential minimization conditions,
\begin{align}
2\la_1v^2+\la_4\La_1^2+\la_5\La_2^2+2\mu_1^2 &= 0,\\
2\la_2\La_1^2+\la_4v^2+\la_6\La_2^2+3\la\La_1\La_2+2\mu_2^2 &= 0,\\
\la\La_1^3+(2\la_3 \La_2^2+\la_5v^2+\la_6\La_1^2+2\mu_3^2)\La_2 &= 0.
\end{align}
Hence, we obtain the mass-squared matrix of the $CP$-even scalars ($S,S_{1,2}$) as
\be M_S^2=\left(\begin{array}{ccc} 2\la_1v^2 & \la_4v\La_1 & \la_5v\La_2 \\
	\la_4v\La_1 & \fr 1 2 (4\la_2\La_1+3\la\La_2)\La_1 & \fr 1 2 (2\la_6\La_1\La_2+3\la\La_1^2) \\
	\la_5v\La_2 & \fr 1 2 (2\la_6\La_1\La_2+3\la\La_1^2) & \fr{1}{2\La_2}(4\la_3\La_2^3-\la\La_1^3)\end{array}\right). \ee

Due to the hierarchy $v\ll \La_1,\La_2$, the mixing matrix $M_S^2$ can be approximately diagonalized using the seesaw method, in which the state $S$ is the lightest and is separated from the heavy states $S_{1,2}$. We obtain the light physical state with its squared mass as 
\be H\simeq S-\ep_1 S_1-\ep_2 S_2 , \hs m_H^2\simeq 2\la_1v^2-(\ep_1 \la_4\La_1+\ep_2 \la_5\La_2)v, \ee 
with the mixing parameters are defined by 
\begin{align} \ep_1 &= \frac{[\la(\la_4\La_1^3+3\la_5\La_1\La_2^2)-2(2\la_3\la_4-\la_5\la_6)\La_2^3]v}{2[3\la^2\La_1^3\La_2+\la(\la_2\La_1^4-3\la_3\La_2^4+3\la_6\La_1^2\La_2^2)-(4\la_2\la_3-\la_6^2)\La_1\La_2^3]},\\
\ep_2 &= \frac{[3\la(\la_4\La_1^2-\la_5\La_2^2)-2(2\la_2\la_5-\la_4\la_6)\La_1\La_2]v\La_2}{2[3\la^2\La_1^3\La_2+\la(\la_2\La_1^4-3\la_3\La_2^4+3\la_6\La_1^2\La_2^2)-(4\la_2\la_3-\la_6^2)\La_1\La_2^3]}. \end{align}
These parameters are tiny because they depend on the ratios $v/\La_{1,2}\ll1$. The heavy states $\mathcal{H}_1\simeq \ep_1 S + S_1$ and $\mathcal{H}_2\simeq \ep_2 S + S_2$ mix with each other via a $2\times 2$ submatrix. Diagonalizing this submatrix, we obtain two physical fields,
\be H_1 = c_\xi \mathcal{H}_1 - s_\xi \mathcal{H}_2, \hs H_2 = s_\xi \mathcal{H}_1 + c_\xi \mathcal{H}_2,\ee
with the corresponding masses,
\bea
m^2_{H_{1,2}} &= &\frac{1}{4\La_2}\left\{4\la_3\La_2^3-\la\La_1^3+(4\la_2\La_1+3\la\La_2)\La_1\La_2\right.\crn
&&\left.\mp\sqrt{[4\la_3\La_2^3-\la\La_1^3-(4\la_2\La_1+3\la\La_2)\La_1\La_2]^2+4(2\la_6\La_2+3\la\La_1)^2\La_1^2\La_2^2}\right\}. \eea
The angle $\xi$, which defines the mixing between $H_1$ and $H_2$, is given by  
\be t_{2\xi} = \fr{2(2\la_6\La_2+3\la\La_1)\La_1\La_2}{4\la_3\La_2^3-\la\La_1^3-(4\la_2\La_1+3\la\La_2)\La_1\La_2}. \ee
We see that the mass of the physical state $H$ is of the order of the electroweak scale $m_{H}\sim \mathcal{O}(v)$. Therefore, we identify $H$ as the 125~GeV Higgs boson discovered at the LHC \cite{ParticleDataGroup:2024cfk}. The heavy states $H_{1,2}$ correspond to new Higgs bosons with masses at the $\La_{1,2}$ scale.

Turning to the $CP$-odd sector, which includes $A,A_{1,2}$, we find a massless eigenstate $A=G_Z$. This state is identified as the Goldstone boson absorbed by the SM $Z$ boson. Conversely, $A_{1,2}$ mix through a $2\times 2$ matrix, yielding two physical fields:
\be G_{Z'} = \frac{\La_1A_1+3\La_2A_2}{\sqrt{\La_1^2+9\La_2^2}},\hs \mathcal{A} = \frac{3\La_2A_1-\La_1A_2}{\sqrt{\La_1^2+9\La_2^2}}.
\ee
Here, $G_{Z'}$ acts as the Goldstone boson associated with the new gauge boson $Z'$, while $\mathcal{A}$ corresponds to a heavy pseudoscalar. The latter has a mass determined by the $\La_{1,2}$ scale,
\be 
m^2_\mathcal{A}= -\frac{\la(\La_1^2+9\La_2^2)\La_1}{2\La_2}, 
\ee
where the parameter $\la<0$ ensures a positive squared mass.

On the other hand,  the dark scalars $R_{\eta,\rho}$ and $I_{\eta,\rho}$ mix within each pair. After diagonalization, we obtain the physical states and their masses as 
\begin{align} R_1 &= c_R R_\eta -s_R R_\rho,\hs R_2=s_R R_\eta +c_R R_\rho,\\
I_1 &= c_I I_\eta -s_I I_\rho,\hs I_2=s_I I_\eta +c_I I_\rho, \end{align} 
\bea && m^2_{R_{1}}\simeq M^2_\eta + \fr{(\sqrt{2}\mu\La_2+\la_{19} v^2)^2}{4(M^2_\eta - M^2_\rho)},\hs m^2_{R_{2}}\simeq M^2_\rho - \fr{(\sqrt{2}\mu\La_2+\la_{19} v^2)^2}{4(M^2_\eta - M^2_\rho)},\label{remass}\\
&& m^2_{I_{1}}\simeq M^2_\eta + \fr{(\sqrt{2}\mu\La_2-\la_{19} v^2)^2}{4(M^2_\eta - M^2_\rho)},\hs m^2_{I_{2}}\simeq M^2_\rho - \fr{(\sqrt{2}\mu\La_2-\la_{19} v^2)^2}{4(M^2_\eta - M^2_\rho)},\label{immass}\eea 
where $M^2_\eta=\mu^2_4+\fr{\la_{11}}{2}v^2+\fr{\la_{15}}{2}\La_1^2+\fr{\la_{17}}{2}\La_2^2$ and $M^2_\rho=\mu^2_5+\fr{\la_{12}}{2}v^2+\fr{\la_{16}}{2}\La_1^2+\fr{\la_{18}}{2}\La_2^2$. The two mixing angles $\theta_{R,I}$ are defined by 
 \be t_{2R,2I}=\fr{\sqrt{2}\mu\La_2\pm\la_{19}v^2}{M^2_{\rho}-M^2_\eta},\label{dsmix}\ee 
which are small because $\mu\La_2\sim \la_{19}v^2\ll M^2_{\eta,\rho}\sim \bar{M}^2$ and $\bar{M}\sim\mathcal{O}(1)$ TeV. 

For the charged scalar sector, $\ph^\pm$ is identified as a massless eigenstate, $G_W^\pm \equiv \ph^\pm$, which acts as the Goldstone boson absorbed by the SM $W^{\pm}$ boson. The remaining fields,  $\eta^\pm$ and $\rho^\pm$, mix through via a $2\times 2$ matrix. After diagonalization, two physical charged Higgs bosons $H_{1}^{\pm}$ and $H_2^{\pm}$ emerge, with masses determined by the $\La_{1,2}$ scale, namely 
\begin{align} H_1^\pm &= c_\theta \eta^\pm -s_\theta \rho^\pm,\hs m^2_{H_1^\pm}\simeq M^2_\eta+\frac{\la_{13}}{2}v^2,\label{chargedmass}\\
H_2^\pm &= s_\theta \eta^\pm +c_\theta \rho^\pm,\hs m^2_{H_2^\pm}\simeq M^2_\rho+\frac{\la_{14}}{2}v^2, \end{align}
assuming $\mu\ll\La_{1,2}$. The small angle $\theta$ represents the mixing between $H_{1}^{\pm}$ and $H_2^{\pm}$ and is given by 
\be t_{2\theta}\simeq\fr{\sqrt{2}\mu\La_2}{M^2_\rho-M^2_\eta}.\ee

We derive the gauge bosons masses from the kinetic terms of the scalar fields, $\mathcal{L}_S=\sum_{S=\chi_1,\chi_2}(D^\mu S)^\dagger (D_\mu S)$. The spontaneous breaking of the $SU(2)_L\times U(1)_Y \times U(1)_X$ symmetry leads to physical masses for the $W^{\pm},Z$ and $Z'$ bosons. The physical states and the corresponding masses of the gauge bosons $W^\pm, A, Z$ are given by
\begin{align} W^\pm &= \frac{1}{\sqrt2}(A_1\mp iA_2),\hs m^2_W = \fr{g^2v^2}{4} ,\\
A &= s_W A_3 + c_W B,\hs m_A=0,\\
Z &= c_W A_3 - s_W B,\hs m^2_Z = \frac{g^2v^2}{4c^2_W},\\
Z' &= C,\hs m^2_{Z'}= 4g^2_Xz^2(\La^2_1+9\La^2_2), \end{align}
where the Weinberg angle is defined as $\tan(\theta_W) = g_Y/g$. For convenience, we also use the shorthand notations $t_W\equiv \tan(\theta_W)$, $s_W\equiv \sin(\theta_W)$, $c_W\equiv \cos(\theta_W)$. We note that in this model the $Z$--$Z'$ mixing is absent because the SM doublet $\ph$ carries zero $X$ charge, while $\chi_{1,2}$ are singlets under $SU(2)_L\otimes U(1)_Y$.

\subsection{Investigating SM-like Higgs coupling properties}
We derive the interactions between $H$ and the SM quarks and leptons by considering the following  Yukawa Lagrangian: 
\bea\mathcal{L}&\supset& h^d_{\al\bet}\bar{q}_{\al L}\ph d_{\bet R}+h^d_{33}\bar{q}_{3L}\ph d_{3R}+\frac{h^d_{\al 3}}{\La_c}\bar{q}_{\al L}\ph\chi_1 d_{3R}+\frac{h^d_{3\bet}}{\La_c}\bar{q}_{3L}\ph\chi_1^* d_{\bet R}\crn
&&+h^u_{\al\bet}\bar{q}_{\al L}\tilde{\ph} u_{\bet R}+h^u_{33}\bar{q}_{3L}\tilde{\ph} u_{3R}+\frac{h^u_{\al 3}}{\La_c}\bar{q}_{\al L}\tilde{\ph}\chi_1 u_{3R}+\frac{h^u_{3\bet}}{\La_c}\bar{q}_{3L}\tilde{\ph}\chi_1^* u_{\bet R}\crn
&&+h^e_{11}\bar{l}_{1L}\ph e_{1R}+h^e_{xy}\bar{l}_{xL}\ph e_{yR}+\frac{h^e_{1y}}{\La_c}\bar{l}_{1L}\ph\chi_2 e_{yR}+\frac{h^e_{x1}}{\La_c}\bar{l}_{xL}\ph\chi_2^* e_{1R}+\mathrm{H.c.},
\eea
where $\tilde{\ph}=i\sigma_2\ph^*$ with $\sigma_2$ being the second Pauli matrix), and $\La_c$ represents the new-physics (cutoff) scale associated with the non-renormalizable interactions. 

We can obtain the interactions of $H$ with quarks and leptons as follows
	\bea
	\mathcal{L}^{H} &\supset& \left\{\bar{q}_{\al L} \fr{[M_q]_{\al\beta}}{v} q_{\bet R}+\bar{q}_{3L} \fr{[M_q]_{33}}{v} d_{3 R} +\bar{q}_{\al L} \fr{[M_q]_{\al3}(1-\ep_1v/\La_1)}{v} q_{3 R}\right.\crn
	&&\left.+\bar{q}_{3 L} \fr{[M_q]_{3\beta}(1-\ep_1v/\La_1)}{v} q_{\beta R}\right\}H  + \left\{\bar{e}_{xL} \fr{[M_e]_{xy}}{v} e_{y R}+\bar{e}_{1L} \fr{[M_e]_{11}}{v} e_{1 R}\right.\crn 
	&&\left. +\bar{e}_{1 L} \fr{[M_e]_{1y}(1-\ep_2v/\La_2)}{v} e_{y R}+\bar{e}_{x L} \fr{[M_e]_{x1}(1-\ep_2v/\La_2)}{v} e_{1 R}\right\}H +\mathrm{H.c.} \crn 
	&\supset& \bar{q}_{aL} \fr{[M_q]_{ab}}{v} q_{b R} + \bar{e}_{aL} \fr{[M_e]_{ab}}{v} e_{b R}+\mathrm{H.c.}, \label{LHyu}
	\eea
	where $q=u,d$ and $[M_{q,e}]_{ij}$ are the mass matrices for quarks and charged leptons, defined by 
	\begin{align}
	[M_q]_{\al\beta} &= -h^q_{\al\beta}\frac{v}{\sqrt2}, \hs [M_q]_{33}=-h^q_{33}\frac{v}{\sqrt2}, \hs \left[M_q\right]_{\al 3} = -h^q_{\al 3}\fr{v\La_1}{2\La_c}, \hs [M_q]_{3\beta}= -h^q_{3\beta}\frac{v\La_1}{2\La_c},\\
	\left[M_e\right]_{11} &= -h^e_{11}\frac{v}{\sqrt2}, \hs [M_e]_{xy} = -h^e_{xy}\frac{v}{\sqrt2},\hs \left[M_e\right]_{1y} = -h^e_{1y}\frac{v\La_2}{2\La_c}, \hs [M_e]_{x1} = -h^e_{x1}\frac{v\La_2}{2\La_c}. 	
\end{align}	 
We see that the matrix elements $[M_q]_{\al 3,\beta 3}$ for quarks and $[M_e]_{1 y,x1}$ for leptons are slightly modified by the new physics contributions $\ep_{1}v/\La_1,\ep_2v/\La_2$. These corrections are suppressed by $v\ll \La_{1,2}$ and $\ep_{1,2}\ll 1$; hence, they can be safely neglected. In general, the matrices $M_{f}$ ($f=q,e$) can be diagonalized by bi-unitary transformations, $\mathrm{M_f} = V_{f_L}^\dag M_f V_{f_R}$, where $V_{f_{L,R}}$ are unitary matrices, relating the gauge eigenstates to the physical states, i.e. $f_{L,R}=V_{f_{L,R}}f'_{L,R}$, and $\mathrm{M_f}$ are diagonal mass matrices. Consequently, the last line in Eq.~(\ref{LHyu}) can be rewritten as
	\be 
	\mathcal{L}^{H}=\bar{e}'_L\frac{\mathrm{M_e}}{v}e'_RH -\bar{d}^\prime_L \frac{\mathrm{M_d}}{v}  d^\prime_R H-\bar{u}^\prime_L \frac{\mathrm{M_u}}{v}  u^\prime_R H +\mathrm{H.c.}
	\label{FCNC1}\ee
We observe that $H$ couples to the fermions $u,d,e$ in the same way as predicted in the SM. 
	
In addition, we would like to emphasize that the scalars $\chi_{1},\chi_2$ have zero hypercharge, $Y_{\chi_{1,2}}=0$, and are singlets under the $SU(2)_L$ symmetry. Therefore, they do not couple to the SM gauge bosons $W^{\pm}$ and $Z$. This implies that $H$ interacts with the gauge bosons $W^{\pm}$ and $Z$ in the same way as predicted in the SM.  
	 
With the presence of the new charged Higgs bosons $H_{1,2}^{\pm}$, the couplings of $H$ to these scalars are obtained as
	\begin{align}
	g_{HH_1^+H_1^-}&=2v(\la_{13}c_{\theta}^2-\la_{14}s_{\theta}^2)+(\ep_1\La_1+\ep_2\La_2)(-\la_{17}c_{\theta}^2+\la_{18}s_{\theta}^2), \crn 
	g_{HH_2^+H_2^-}&=2v(\la_{14}c_{\theta}^2-\la_{13}s_{\theta}^2)-(\ep_1\La_1+\ep_2\La_2)(\la_{17}s_{\theta}^2+\la_{18}c_{\theta}^2)-\sqrt{2}\ep_2\mu s_{\theta}c_{\theta}.\label{gS}  
	\end{align}
	
\section{\label{m3} Physical constraints derived from phenomenology of SM-like Higgs boson $H$}
To test the consistency of our model with LHC data, we employ the $\ka$ parameterization  \cite{ParticleDataGroup:2024cfk} for the Higgs boson couplings. The parameters $\ka_i$ with $i=l,u,d,W,Z$ are defined as the ratios $\ka_i=g_i/g^{\text{SM}}_i$, which serve as probes of possible new-physics contributions. For the couplings of $H$ to fermions, using Eq.~(\ref{FCNC1}), we obtain $\ka_f=g_f/g_f^{\text{SM}}=1$, which is consistent with the current ATLAS and CMS Run II experimental constraints \cite{ParticleDataGroup:2024cfk}. We also remark that with the predicted values $\ka_f=1$, the parameter for the effective Higgs coupling to gluons $\ka_g$, which depends on the fermionic couplings, becomes $\ka_g^2=1.042\ka_t^2-0.040\ka_b\ka_t+0.002\ka_b^2-0.005\ka_t\ka_c+0.0005\ka_b\ka_c+0.00002\ka_c^2=1$, which is also consistent with the experimental measurements \cite{ParticleDataGroup:2024cfk}. Additionally, the modified parameters describing the couplings of $H$ to the SM gauge bosons $W^{\pm}$ and $Z$ remain unchanged and are equal to unity, namely $\ka_W=g_{HW^+W^-}/g^{\text{SM}}_{HW^+W^-}=1$, $\ka_Z=g_{HZZ}/g^{\text{SM}}_{HZZ}=1$. This is because the scalars $\chi_{1,2}$ (which contribute to $H$ through the mixing parameters $\ep_{1,2}$) do not interact with the gauge bosons $W^{\pm}$ and $Z$, as they carry zero hypercharge and are singlets under the $SU(2)_L$ symmetry, as mentioned above. 

We can then determine the decay width $\Gamma(H\to \gamma\gamma)$ as
	\be 
	\Ga(H\to \ga\ga)=\fr{G_F\al^2m_{H}^3}{128\sqrt{2}\pi^3}\left|N_CQ_t^2\ka_tA_f(x_t)+Q_W^2\ka_WA_W(x_W)+\sum_{S}\fr{m_{W}^2}{m_S^2}Q_S^2\ka_SA_S(x_S)\right|^2,
	\ee 
where the color factor is $N_C=3$, the top-quark electric charge $Q_t=2/3$, and $Q_S$ denotes the electric charge of the charged scalars $S=\{H_1^{\pm},H_2^{\pm}\}$. The parameters $\ka_S$ are defined as $\ka_S=\{g^{HH_1^+H_1^-}/g^{\text{SM}}_{S},g^{HH_2^+H_2^-}/g^{\text{SM}}_{S}\}$ with $g^{\text{SM}}_{S}=-g^2v/2$. The loop functions $A_{f,W,S}(x_{f,W,S})$ with  $x_{t,W,S}=m_{H}^2/(4m_{t,W,S}^2)<1$ are given by  
	\begin{align}
	A_f(x_f)&=\fr{2}{x^2_f}[x_f+(x_f-1)\arcsin^2\sqrt{x_f}],\crn
	A_S(x_S)&=\fr{-(x_S-\arcsin^2\sqrt{x_S})}{x_S^2},\crn 
	A_W(x_W)&=\fr{-[2x_W^2+3x_W+3(2x_W-1)\arcsin^2\sqrt{x_W}]}{x_W^2}.
	\end{align} 
The parameter $\ka_{\ga}$ is then defined as
	\be 
	\ka_{\ga}^2=\fr{\Ga(H\to \ga\ga)}{\Ga(H\to \ga\ga)_{\text{SM}}},
	\ee 	
where $\Ga(H\to \ga\ga)_{\text{SM}}$ denotes the SM prediction. We compare the predicted values $\ka_{\ga}$ with the latest LHC Run-II data, namely $\ka^{\text{ATLAS}}_{\ga}=1.01\pm 0.06$ and $\ka^{\text{CMS}}_{\ga}=1.1\pm 0.08$~\cite{ParticleDataGroup:2024cfk}. We emphasize that $\ka_{S}$ can also affect another modified parameter, $\ka_{Z\ga}$. However, the current experimental uncertainties reported by both ATLAS and CMS for this channel are still significantly larger than those for the $\ka_{\ga}$ channel. Therefore, we do not consider $\ka_{Z\ga}$ in the present work.

\subsection{Numerical study}
For the numerical analysis, the following input values from the Particle Data Group (PDG)~\cite{ParticleDataGroup:2024cfk} are used: 
	\begin{align}
	 m_t &\simeq 173 \ \text{GeV}, \hs m_{H}\simeq 125  \ \text{GeV}, \hs m_{W}=80.385  \ \text{GeV}, \hs m_{Z}=91.67  \ \text{GeV}, \crn 
	s_W^2 &\simeq 0.231, \hs \al_{\text{em}}\simeq 1/128, \hs \al_s(m_Z)\simeq 0.118.
	\end{align}  
	For simplicity, the quartic couplings  between the SM-like doublet $\phi$ and the odd doublets $\eta$ (or $\rho$) with different structure are assumed to be equal, i.e. $\la_{11}=\la_{13}$ and $\la_{12}=\la_{14}$. This assumption is also applied for the quartic couplings between the singlets $\chi_1,\chi_2$ and the odd doublets $\eta,\rho$, namely $\la_{15}=\la_{17}$ and $\la_{16}=\la_{18}$. In addition, we impose the following constraints derived from previous work:
	\begin{itemize}
		\item Constraint from quark flavor-violating observables: 57.09 TeV$\leq 2\sqrt{\La_1^2+9\La_2^2}\leq $ 78.92 TeV. 
		\item Small mixing angle in the charged Higgs sector: $\theta\sim 0.5\times \arctan{v^2/|m^2_{H_{1^{\pm}}}-m^2_{H_2^{\pm}}|}<\pi/16$.
		\item Constraints from charged-lepton flavor–violating observables: $m_{H_{1,2}^{\pm}}\in[0.5,1.2]$ TeV.
		\end{itemize}
		
\begin{figure}[H]	
		\centering
		\begin{tabular}{cc}		
			\includegraphics[width=7.5cm]{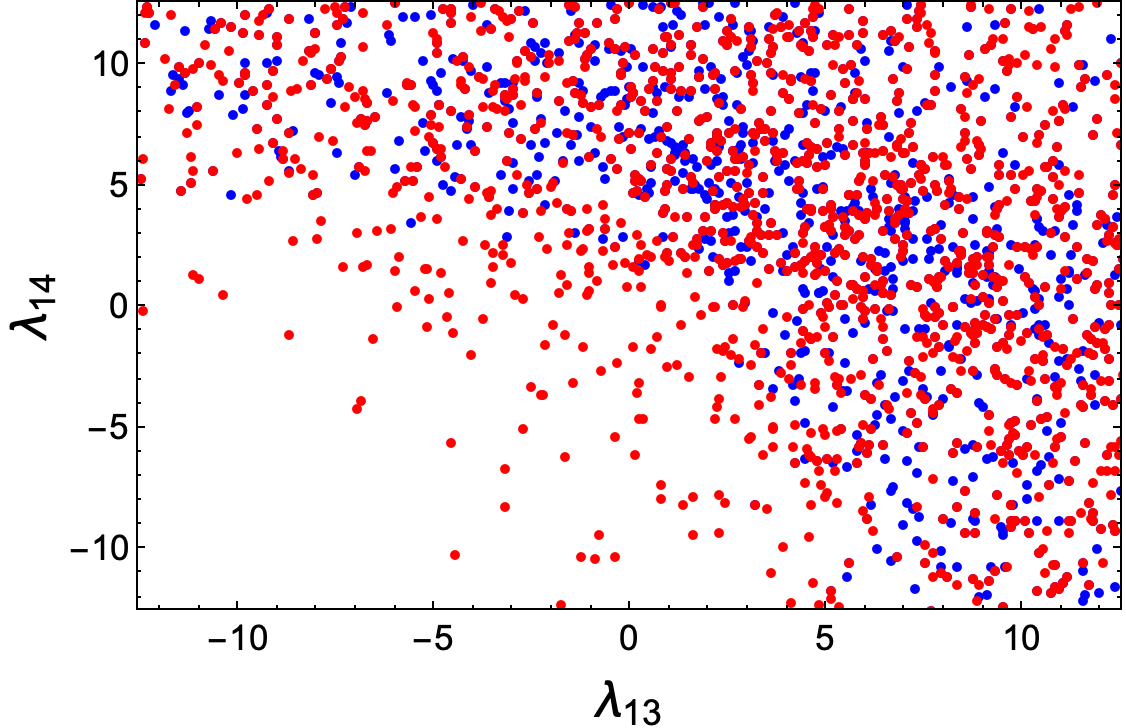}& 		
			\includegraphics[width=7.5cm]{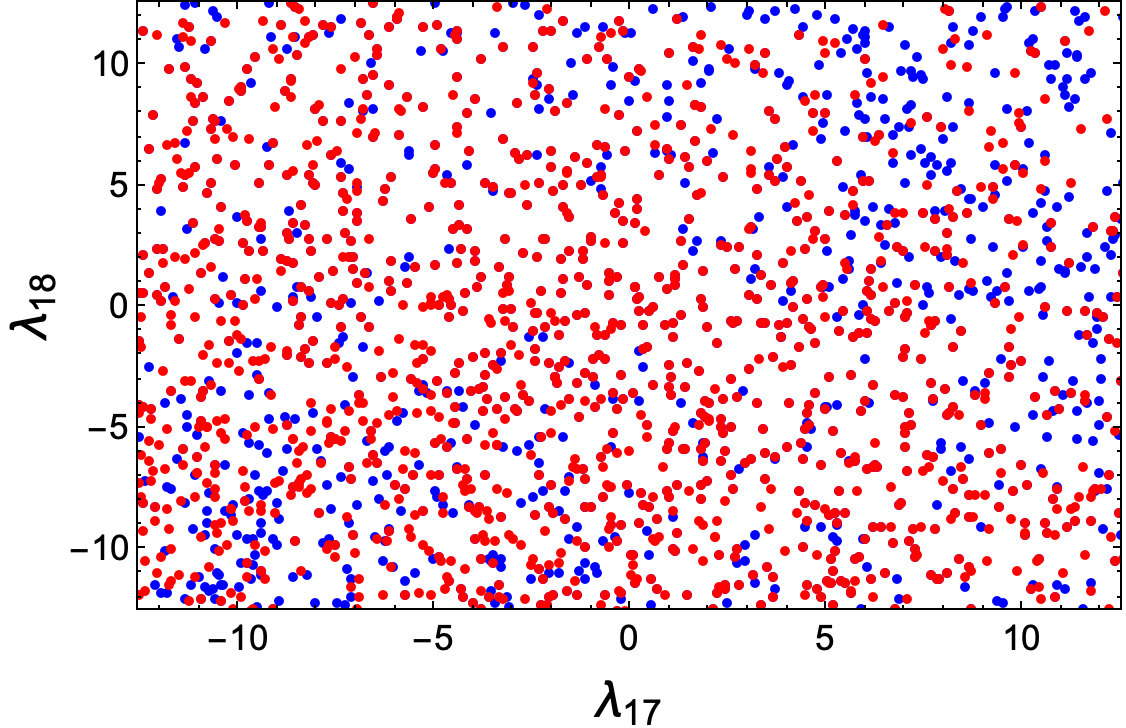}\\
		\end{tabular}	
		\caption{\label{la-limit}Correlations between the scalar couplings constrained by the $1\sigma$ ATLAS and CMS bounds on $\ka_{\ga}$. Left panel: $\la_{13}$ versus $\la_{14}$. Right panel: $\la_{17}$ versus $\la_{18}$. The blue and red points correspond to the mixing parameters $\ep_{1,2}=10^{-3}$ and $\ep_{1,2}=10^{-2}$, respectively.}	
	\end{figure}

Since $\theta$ is small, Eq.~(\ref{gS}) can be written in a simpler form as 
	\bea 
	 g_{HH_1^+H_1^-}\simeq 2v\la_{13}-(\ep_1\La_1+\ep_2\La_2)\la_{17}, \hs 
	g_{HH_2^+H_2^-}\simeq 2v\la_{14}-(\ep_1\La_1+\ep_2\La_2)\la_{18}.\label{gSp}  
	\eea
	The quartic couplings $\la_{13},\la_{14},\la_{17},\la_{18}$ are constrained by the perturbativity condition, i.e. $|\la_{ij}|<4\pi$. The mixing parameters $\ep_1$ and $\ep_2$ are suppressed at the order $\mathcal{O}(v/\La_{1,2})\ll 1$; therefore, we fix their values to either $10^{-2}$ or  $10^{-3}$. With this setup, the modified parameter $\ka_{\ga}$ depends on the new physics scales $\La_1,\La_2$, and several scalar couplings $\la_{13},\la_{14},\la_{17},\la_{18}$. The panels in Fig .\ref{la-limit} show the correlations between the scalar couplings that satisfy the $1\sigma$ experimental constraints on the modified parameter $\ka_{\ga}$ reported by both the ATLAS and CMS collaborations.

In the left panel of Fig. \ref{la-limit}, a strong correlation between the quartic couplings $\la_{13}$ and $\la_{14}$can be observed: $\la_{13}$ increases as $\la_{14}$ decreases, and vice versa. For mixing parameters $\ep_{1,2}=10^{-3}$, the allowed values of $\la_{13,14}$ mainly lie in the upper-right region of the panel and satisfy the relation $\la_{13}+\la_{14}\geq 5-4\pi$. When $\ep_{1,2}=10^{-2}$, the allowed region becomes larger and approximately follows $\la_{13}+\la_{14}\geq -4\pi$. This behavior indicates that the allowed range of $\la_{13}-\la_{14}$ expands as the mixing parameters $\ep_{1,2}$ increase. The reason is that larger values of $\ep_{1,2}$ enhance the contributions induced by the mixing between the SM-like Higgs boson $H$ and the heavy Higgs states $H_{1,2}$, which can become comparable to the non-mixing contributions. As shown in Eq .(\ref{gSp}), these two terms tend to partially cancel each other, thereby suppressing the new-physics contribution to $\ka_{\ga}$ and enlarging the allowed parameter region. In contrast, the right panel of Fig.~\ref{la-limit} shows that the entire parameter space of the scalar couplings $\la_{17}$ and $\la_{18}$ satisfies the $1\sigma$ experimental bounds on $\ka_{\ga}$. This indicates that the effects of $\la_{17,18}$ are much less significant than those of $\la_{13,14}$ in determining the value of $\ka_{\ga}$. 
	
\begin{figure}[H]	
		\centering
		\begin{tabular}{cc}		
			\includegraphics[width=8.5cm]{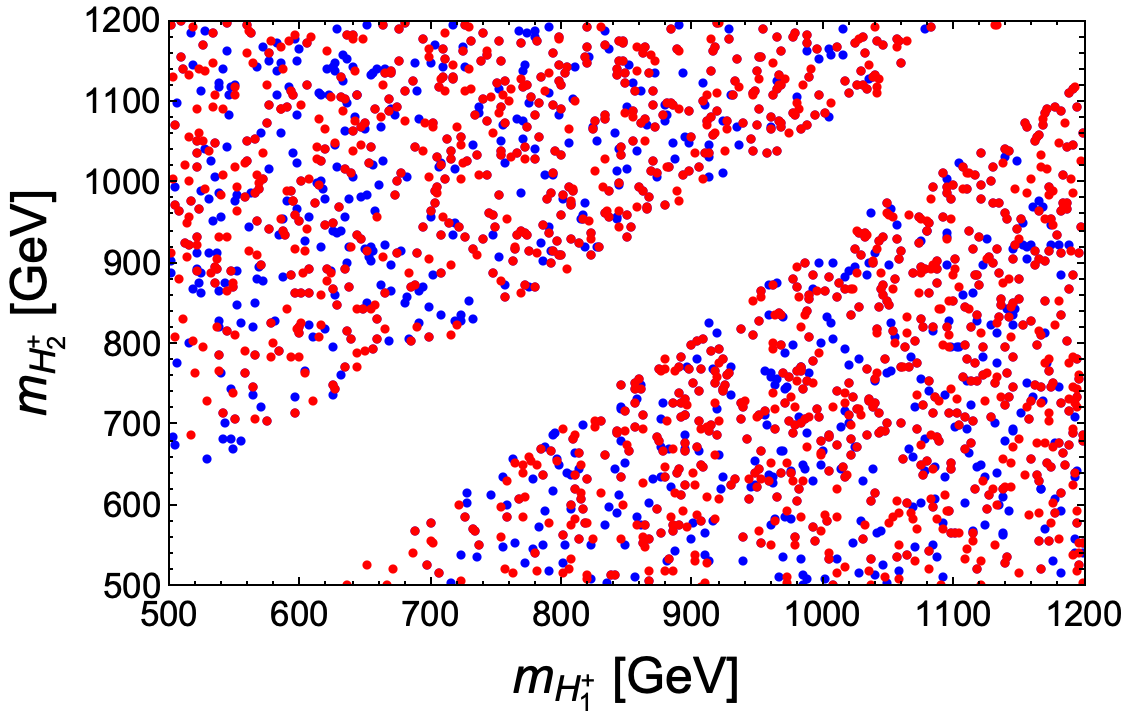}
		\end{tabular}	
		\caption{\label{mhc-limit}Correlation between the charged Higgs masses $m_{H_{1}^{\pm}}$ and $m_{H_{2}^{\pm}}$ satisfying the $1\sigma$ ATLAS and CMS constraints on $\ka_{\ga}$. The blue and red dots correspond to the mixing parameters $\ep_{1,2}=10^{-3}$ and $\ep_{1,2}=10^{-2}$, respectively.}			\end{figure}

Turning to Fig~.\ref{mhc-limit}, we obtain stronger constraints on the charged Higgs masses $m_{H_{1}^{\pm}}$ and $m_{H_2^{\pm}}$ compared to those reported in Ref.~\cite{VanLoi:2024ptt}. In particular, the lower bounds are found to be $m_{H_1^{\pm}}\geq 700$ GeV and $m_{H_2^{\pm}}\geq 660$ GeV for $\ep_{1,2}=10^{-3}$, while for $\ep_{1,2}=10^{-2}$ they become $m_{H_1^{\pm}}\geq 650$ GeV and $m_{H_2^{\pm}}\geq 680$ GeV. 
	
\section{\label{m4}Conclusion}
In this work, we have systematically investigated the phenomenology of the light Higgs boson $H$ in a flavor-dependent $U(1)$ Abelian gauge model. We find that $H$ couples to all SM particles, including quarks and charged leptons $f=u,d,l$, as well as the gauge bosons $W^{\pm}$ and $Z$, in the same way as predicted in the SM. We have also studied the modified parameter $\ka_{\ga}$. In our model, this parameter can receive additional contributions from the interactions between the SM Higgs doublet $\phi$ and the new doublets $\eta,\rho$, as well as from the mixing between the SM-like Higgs $H$ and the heavy Higgs bosons $H_{1,2}$. Our results show that the predicted values of $\ka_{\ga}$ are consistent with both the ATLAS and CMS measurements at the $1\sigma$ level, while simultaneously satisfying the constraints from collider, flavor, and dark matter studies obtained in Ref.~\cite{VanLoi:2024ptt}. We find that the quartic couplings $\la_{17}$ and $\la_{18}$ have negligible effects on $\ka_{\ga}$, whereas $\la_{13}$ and $\la_{14}$ play a crucial role. In particular, the allowed parameter space satisfies the relations $\la_{13}+\la_{14}\geq 5-4\pi$ for $\ep_{1,2}=10^{-3}$ and $\la_{13}+\la_{14}\geq -4\pi$ for $\ep_{1,2}=10^{-2}$. Furthermore, stronger bounds on the charged Higgs masses are obtained, namely $m_{H_1^{\pm}}\geq 700$ GeV and $m_{H_2^{\pm}}\geq 660$ GeV for $\ep_{1,2}=10^{-3}$, while $m_{H_1^{\pm}}\geq 650$ GeV and $m_{H_2^{\pm}}\geq 680$ GeV for $\ep_{1,2}=10^{-2}$. 
	
\section*{Acknowledgements}
N.T.Duy is supported by the Vietnam Academy of Science and Technology, under Grant No. CBCLCA.03/25-27.
	
\bibliographystyle{JHEP}

\bibliography{HiggsU1}

\end{document}